\documentclass[useAMS,usenatbib]{mn2e}
\usepackage{graphicx}

\newcommand{\etal}{{et~al.}}
\newcommand{\lsim}{\,\lower2truept\hbox{${<\atop\hbox{\raise4truept\hbox{$\sim$}}}$}\,}
\newcommand{\gsim}{\,\lower2truept\hbox{${>\atop\hbox{\raise4truept\hbox{$\sim$}}}$}\,}
\newcommand{\beq}{\begin{equation}}
\newcommand{\eeq}{\end{equation}}
\newcommand{\COBE}{$COBE$-DMR}
\newcommand{\fstica}{{\sc{FastICA}}}

\newcommand{\vect}[1]{{\mathbfit{#1}}}
\newcommand{\vev}[1]{\langle#1\rangle}
\newcommand{\ha}{H$\alpha$}
\newcommand{\um}{$\mu$m}

\newcommand{\qrms}{$Q_{rms-PS}$}
\newcommand{\n}{$n$}

\begin{document}

\title[Astrophysical components separation of \COBE\ data with \fstica]
{Astrophysical components separation of \COBE\ 4yr data with \fstica}

\author[D. Maino et al.]
{D. Maino,$^{1}$\thanks{davide.maino@mi.infn.it}
A. J. Banday$^{2}$, C. Baccigalupi$^3$, 
F. Perrotta$^3$, K. M. G\'{o}rski,$^{4,5}$\thanks{\emph{Current 
address: Jet Propulsion Laboratory, California Institute of 
Technology, 4800 Oak Grove Drive, Pasadena CA 91109, USA.}} \\
$^{1}$ Dipartimento di Fisica, Universit\`a di Milano, 
Via Celoria 16, 20133, Milano, Italy.\\
$^{2}$ Max-Planck Institute f{\"u}r Astrophysik, 
Karl-Schwarzschildstrasse 1, 85741, Garching bei 
M\"{u}nchen, Germany.\\
$^{3}$ SISSA/ISAS, Astrophysics Sector, Via 
Beirut 4, 34014, Trieste, Italy.\\
$^{4}$ ESO, Karl-Schwarzschildstrasse 2, 85740, Garching bei
M\"{u}nchen, Germany.\\
$^{5}$ Warsaw University Observatory, Aleje Ujazdowskie 4, 
00-478 Warszawa, Poland.}

\date{Receiverd
**insert**; Accepted **insert**}

\pagerange{\pageref{firstpage}--\pageref{lastpage}}
\pubyear{2003}

\maketitle

\label{firstpage}

\begin{abstract}
We present an application of the fast Independent Component Analysis
\fstica\ method to the \COBE\ 4yr data. Although the signal-to-noise ratio in
the \COBE\ data is typically $\sim 1$, the approach is able to extract the CMB signal 
with high confidence when working at high galactic latitudes.
However, the foreground emission components
have too low a $S/N$ ratio to be reconstructed by this method
(moreover, the number of components which can be reconstructed is
directly limited by the number of input channels).

The reconstructed CMB map shows the expected frequency scaling of the
CMB. 
We fit the resulting CMB component for the rms quadrupole normalisation \qrms\ 
and primordial spectral index \n\ and find results in excellent
agreement with those derived from the minimum-noise combination of the 90 and 53~GHz
DMR channels without galactic emission correction.

We extend the analysis by including additional channels (priors) 
such as the Haslam map of radio emission at 408~MHz and
the DIRBE 140\um\ map of galactic infra-red emission. 
Subsequently, the \fstica\ algorithm is able to both detect galactic 
foreground emission and separate it from the dominant 
CMB signal. Fitting the resulting CMB component for
\qrms\ and \n\ we find good agreement with the results from G\'orski
\etal~(1996) in which the galactic emission has been taken into account by 
subtracting that part of the DMR signal observed to be
correlated with these galactic template maps.
\fstica\ is therefore able to extract 
foreground emission from the DMR data and recover a ``clean'' CMB 
component. 

We further investigate the ability of \fstica\ to evaluate the extent of foreground
contamination in the \COBE\ data. We include an all-sky \ha\ survey 
(Dickinson, Davies \& Davis 2003) to determine a reliable free-free
template of the diffuse interstellar medium which we 
use in conjunction with the previously described synchrotron and dust templates. 
The derived frequency scalings of the recovered foregrounds is
consistent with previous correlation studies (Banday \etal\ 2003).
In particular  we find that, after subtraction
of the thermal dust emission predicted by the Finkbeiner, Davis \& Schlegel (1999)
model 7, this component is the dominant foreground emission at 31.5~GHz.
This indicates the presence of an anomalous dust correlated component which
is well fitted by a power law spectral shape $\nu^{-\beta}$ with
$\beta \sim 2.5$ in agreement with Banday \etal\ (2003).

\end{abstract}


\section{Introduction}
\label{intro}

Current and future CMB space missions such as $WMAP$ 
(launched in June 2001, Bennett \etal~1996) and {\sc Planck} 
(scheduled for launch in February 2007, Tauber 2000) 
will map the microwave
sky emission over the entire sky with an unprecedented combination of
sensitivity and angular resolution. In the meantime a large variety of ground-based
and balloon-borne experiments will provide accurate data over various regions of the sky.

The main goal of such experiments is to produce a map 
cleaned of any contributions from other sources between us and the Last Scattering 
Surface and thus containing only genuine CMB anisotropy. From such
maps the angular power spectrum is evaluated and used to determine
cosmological parameters with high accuracy and thereby probe various scenarios for 
structure formation.
Potential foreground contaminants include emission from our Galaxy (mainly 
synchrotron, free-free and dust emission),
compact galactic and extra-galactic sources, and the thermal and 
kinematic Sunyaev-Zel'dovich effect from clusters of galaxies. 
In order to exploit the cosmological information encoded into 
the CMB angular power spectrum, it
is crucial to be able to identify and remove such signal components with 
high accuracy and reliability.

Much recent work has been performed in this area by several authors 
(de Oliveira-Costa \& Tegmark 1999, Tenorio \etal 1999, Hobson \etal~1998, 1999,
Stolyarov \etal~2001, Prunet \etal~2001)
employing a range of techniques from the classical Wiener Filter
to maximum entropy methods. These algorithms are referred to as
``non-blind'' in the sense that they require some a-priori information
on the signal to be separated
(e.g. spatial templates and frequency dependences of the underlying
components, although see Barreiro \etal~2003 for a variation on this approach
using a maximum entropy based technique). 
Recently a blind approach has been proposed (Baccigalupi \etal~2001,
Maino \etal~2002) and applied to simulated sky maps similar to those that 
{\sc Planck} will produce.
The approach appears to be very promising in that it is fast and does not need priors 
about the underlying signals (at least for high sensitivity missions
such as $WMAP$ or {\sc Planck}). It is therefore possible to use such 
blind algorithms to obtain priors that can be fed into classical Bayesian 
separation algorithms. 

However, to-date these blind algorithms have not been applied to real CMB data to 
assess and validate their reliability in a real world situation. We present here the first 
application of \fstica\ to real CMB measurements from the \COBE\ 4yr data.

This paper is organised as follows: 
in Section~\ref{fastica} we briefly present the component separation problem and
the main simplifications required for the application of \fstica. In 
Section~\ref{algo} and \ref{DMR} we summarise the basics of the \fstica\ algorithm 
and its application to \COBE\ data. Section~\ref{noiseimpact} shows the 
impact of the noise amplitude and distribution on \fstica\ outputs which
drives the choice of optimal separation technique.
The main CMB-related results are presented in Section~\ref{results-cmb} 
while Section~\ref{results-fore} considers the implications for foreground emission.
A critical discussion follows in Section~\ref{conclusion}.

\section{Component Separation Problem}
\label{fastica}

Let us suppose that the observed sky radiation is the superposition of $N$ different
physical processes and its frequency and spatial dependences can
be factorized into two separated terms:
\beq
\tilde{x}(\vect{r},\nu) = \sum_{j=1}^N \bar{s}_{j}(\vect{r})f_{j}(\nu)\, .
\eeq

This signal is, in general, observed by an experiment with $M$-frequencies 
through an optical system, whose beam pattern is in general modelled at 
each frequency as a shift-invariant point spread function 
$B(\vect{r},\nu)$. Let us further suppose that $B(\vect{r},\nu)$
is frequency-independent at least within each frequency bandwidth $t(\nu)$.
In addition any real experiment adds some instrumental noise to the
 output $\epsilon_\nu(\vect{r})$.
Following our assumptions the observed signal at a frequency $\nu$ is given by:
\begin{eqnarray}
x_\nu(\vect{r}) & = & \sum_{j=1}^{N} B_\nu(\vect{r}) \ast \bar{s}_j(\vect{r}) \cdot
\int t_{\nu}(\nu ') f_j(\nu ') d\nu ' +
\epsilon_\nu(\vect{r}) \nonumber \\
	& = & B_\nu(\vect{r}) \ast \sum_{j=1}^{N} a_{\nu j} 
	\bar{s}_j(\vect{r}) + \epsilon_\nu(\vect{r}) \\
\label{icadm}
\nonumber
\end{eqnarray}
where $*$ denotes convolution, and
\beq
a_{\nu j} = \int_{}^{} t_{\nu}(\nu ') f_{j}(\nu ') d\nu ' .
\eeq

Our data model can be further simplified assuming that the radiation pattern
of the telescope is frequency independent i.e. $B_\nu(\vect{r}) = B(\vect{r})$.
In this case, Eq.~(2) can be written in vector form as:
\beq
{\bf{x}}(\vect{r}) = {\bf{A}} {\bar{\bf{s}}}(\vect{r}) * B(\vect{r}) +
{\bmath{\epsilon}}(\vect{r}) = {\bf{A}} {\bf{s}}(\vect{r}) + {\bmath{\epsilon}}(\vect{r})\, ,
\eeq
where each component, $s_{j}$, of the vector ${\bf{s}}$ is the
corresponding source function convolved with the $B$ beam pattern.
The matrix ${\bf{A}}$ is the mixing matrix with elements given by
the $a_{\nu j}$ coefficients.

We present this derivation in order to stress data model assumptions.
It is worth noting that one or more of these 
might be not justified in real cases. For instance both the $WMAP$ and {\sc 
Planck} experiments will observe the sky radiation through a telescope 
with multi-frequency receivers and the resulting beam pattern is strongly 
frequency dependent. This is up to now one of the main limitation of the 
\fstica\ approach to astrophysical components separation and forced Maino 
\etal~(2002) to further convolve simulated {\sc Planck} sky maps in order 
to obtain similar beam functions. Furthermore the noise term 
${\bf{\epsilon}(\vect{r})}$ is usually
assumed to be additive, signal-independent, white, Gaussian, stationary and
uniformly distributed on the sky. These are additional strong assumptions since 
the noise spectrum of a real experiment may contain a low-frequency tail due to the 
so-called $1/f$ noise and different scanning (observing) strategies
which distribute the integration time on the
sky in a non-uniform way.

\subsection{\fstica\ algorithm}
\label{algo}

We briefly summarise here the \fstica\ algorithm. The problem of obtaining both
the mixing matrix ${\bf{A}}$ and the signals ${\bf{s}}$ from observed data
${\bf{x}}$ is unsolvable if additional information is not provided. 
The ICA approach assumes that
\begin{itemize}
\item the signals ${\bf{s}}$ are independent random processes on the map domain;
\item all the signal, but at most one, have non-Gaussian distribution.
\end{itemize}
The strategy we exploit here is described in detail in Hyv\"arinen \& Oja (1997) and
Hyv\"arinen (1999) and its application in an astrophysical context 
can be found in Maino \etal~(2002):
independent components are extracted maximising a suitable measure of non-Gaussianity.
Indeed the central limit theorem states that a variable which is a mixture of independent 
variables is ``more Gaussian'' than the original ones. Therefore we have to find a 
transformation such that the Gaussianity of the variables is reduced: this is equivalent to
finding a set of transformed variables that are ``more independent'' than the original ones.
Furthermore since our data model is a noisy, we have to define a measure of 
non-Gaussianity that is robust against noise, this is the the so-called neg-entropy.

Approximations to neg-entropy have been given by Hyv\"arinen \& Oja (2000) 
and Hyv\"arinen (1999) and, if the noise has the properties assumed previously
and its covariance matrix is known, the Gaussian moments of the 
transformed  variables ${\bf{y}}$ = ${\bf{W}}{\bf{x}}$ are shown to be 
robust estimates  of the desired functions. Here the matrix ${\bf{W}}$ is the
separation matrix such that the ${\bf{y}}$ components are in fact independent.

The algorithm needs
a preprocessing step (Hyv\"arinen 1999) in which the input maps are ``quasi whitened''.
This reduces the number of unknowns in the problem. Let us assume that
we know the covariance matrix ${\mathbf{\Sigma}}$ of the instrumental noise; at 
each frequency, 
the mean value is removed from the data (the offsets of each independent  
component can be recovered at the end of the separation process) 
and their covariance matrix ${\mathbf{C}}$ is evaluated by computing the 
following expectation value:
\beq
{\mathbf{C}} = \vev{\mathbf{x}\mathbf{x}^T} \, .
\eeq
A modified covariance matrix and quasi-whitened data sets are respectively
given by:
\beq
\hat{\mathbf{\Sigma}} = (\mathbf{C} - \mathbf{\Sigma})^{-1/2} {\mathbf{\Sigma}}
(\mathbf{C} - \mathbf{\Sigma})^{-1/2}\, ,
\eeq
\beq
\hat{\mathbf{x}} = (\mathbf{C} - \mathbf{\Sigma})^{-1/2} \mathbf{x} \, .
\eeq

The separation matrix ${\mathbf{W}}$ is estimated row by row i.e. one component
at a time. Let ${\bf{w}}$ be a $M$-vector such that ${\bf{w}}^T{\hat{\bf{x}}}$
gives one component of the transformed vector ${\bf{y}}$
(${\bf{w}}^T$ is a row of the separation matrix ${\bf{W}}$).
In order to find an estimation of the transformed vector ${\bf{y}}$
that is robust against noise, the following iterative algorithm is applied, together
with a convergence criterion:
\begin{enumerate}
\item choose an initial vector ${\bf{w}}$
\item update it by means of:
$$
{\bf{w}}_{\rm new} = \vev{ \hat{\bf{x}}g({\bf{w}}^T\hat{\bf{x}})} - 
(I + \hat{\bf{\Sigma}})\vev{g'({\bf{w}}^T\hat{\bf{x}})}
$$
where $g$ is a regular non-quadratic function i.e.
$g(u) = u^3$, $g(u) = {\rm tanh}(u)$ and $g(u) = u~ {\rm exp}(-u^2)$.
\item normalise ${\bf{w}_{\rm new}}$ to be a unit vector
\item compare ${\bf{w}_{\rm new}}$ with the old computed value; if not
converged come back to (ii); if converged, begin another row.
\end{enumerate}

This procedure maximises the non-Gaussianity of the component ${\bf{w}}^T{\hat{\bf{x}}}$.
If $k$ rows of the matrix ${\bf{W}}$ are found at a given time, the $k+1$ row is searched
for in a sub-space orthogonal to the first $k$ rows. To this purpose a orthogonalisation
procedure (e.g. by mean of Gram-Schmidt rule) is inserted between point (ii) and (iii).
Once the separation matrix ${\bf{W}}$ is obtained the underlying components are derived
by using:
\beq
{\bf{x}} = {\bf{W}^{-1}}{\bf{y}}
\label{recon}
\eeq
This equation allows us to derive the frequency scalings for each independent component:
the scaling between $\nu$ and $\nu '$ of the $j^{\rm th}$ component is given by the ratio
of ${W}^{-1}_{\nu j}/{W}^{-1}_{\nu ' j}$. 

It is also possible to recover signal-to-noise ratio for the reconstructed components.
Since the noise covariance ${\bf{\Sigma}}$ is supposed to be known, 
noise constrained  realization ${\bf{n_x}}$ for each frequency channels 
can be built. Once we have the separation matrix ${\bf{W}}$, the noise 
realization in \fstica, outputs are given
by ${\bf{W}}{\bf{n_x}}$. Noise is transformed like signals and an estimation of the
noise in the reconstructed component is achieved. This is quite useful: suppose that
we have $M$ frequency channels but only $N<M$ components have to be separated. Using
all $M$ channels we can reconstruct $M$ components and the ``fictitious'' ones will
be characterised by a signal-to-noise ratio lower than unity.

\subsection{Application to \COBE\ 4yr data}
\label{DMR}

The DMR experiment has observed full-sky microwave emission 
at three frequencies (31.5, 53 and 90~GHz). 
At each frequency, two separate channels (denoted A and B) 
measured the temperature difference between two horn
antennas of angular resolution $\sim 7^\circ$ pointing 
in directions separated by $60^\circ$ on the sky. 
This differencing information was then used to reconstruct
full-sky maps of the microwave sky.

The instrumental noise is almost 
white and Gaussian but, due to the scanning strategy, it is not 
uniformly distributed on the sky: variations by a factor of $\sim 4$ is typically 
observed at 90 and 53~GHz while factor of $\sim 6$ is found at 31.5 GHz.
This larger factor reflects the fact that part of the original 31~GHz data have been
discarded due to a well known systematic effect (Kogut \etal~1996). In this
way the 31.5~GHz channels shows a noise distribution on the sky which
is substantially different from that at 53 and 90~GHz.
Furthermore the $S/N$ ratio is quite poor:
for 10$^\circ$ effective resolution smoothed maps this is
$\sim 0.5, 1.5$ and 1 when combining A and B channels at 31.5, 53 and 90~GHz 
respectively, and becomes $\sim 2$ when combining together all the frequencies.
However, no extra-smoothing is required 
before applying \fstica\ since the angular resolution is the same for each channel.

We have applied the \fstica\ algorithm to the \COBE\ 4yr data in HEALPix  
format (G\'orski \etal~1999) with a resolution parameter 
$N_{\rm side}$ = 32 corresponding to 12288 pixels in the sky with
size $\sim 1.83^\circ$. 
The A and B radiometers at each frequency are combined to obtain a sum
map (A+B)/2.

Maino \etal (2002) have shown that the optimal CMB reconstruction, both in terms of 
frequency scaling and offset normalisation, can be achieved by exploiting
data at high galactic latitudes. 
We use the extended custom galactic cut (Banday \etal~1997) to select
regions useful for this CMB reconstruction. Before applying the
code the best fit monopole, dipole and quadrupole -- 
computed using only those pixels surviving the Galactic cut --
are subtracted from the maps. \fstica\ ``naturally'' removes
a monopole from the input sky maps but does not do the same for dipole and
quadrupole distributions. These modes are removed since we expect that structures 
on the largest angular scales are contaminated most significantly 
by a galactic emission which is characterised by a steeply falling power
spectrum. Since one figure of merit for the separation performed by 
\fstica\ is the frequency scaling, this procedure minimises the
contamination of the derived CMB scaling by foreground emission.

We work with all the non-quadratic functions described before:
$g(u) = u^3$, $g(u) = {\rm tanh}(u)$ and $g(u)=u~ {\rm exp}(-u^2)$ identified
by {\em p}, {\em t} and {\em g}, respectively.

Finally, we make use of the $S/N$ ratio and frequency scaling of 
each reconstructed component to assess its physical nature.

\section{Impact of noise distribution and amplitude}
\label{noiseimpact}

\begin{figure*}
\begin{center}
\includegraphics[width=3.in,height=6.in,angle=90]{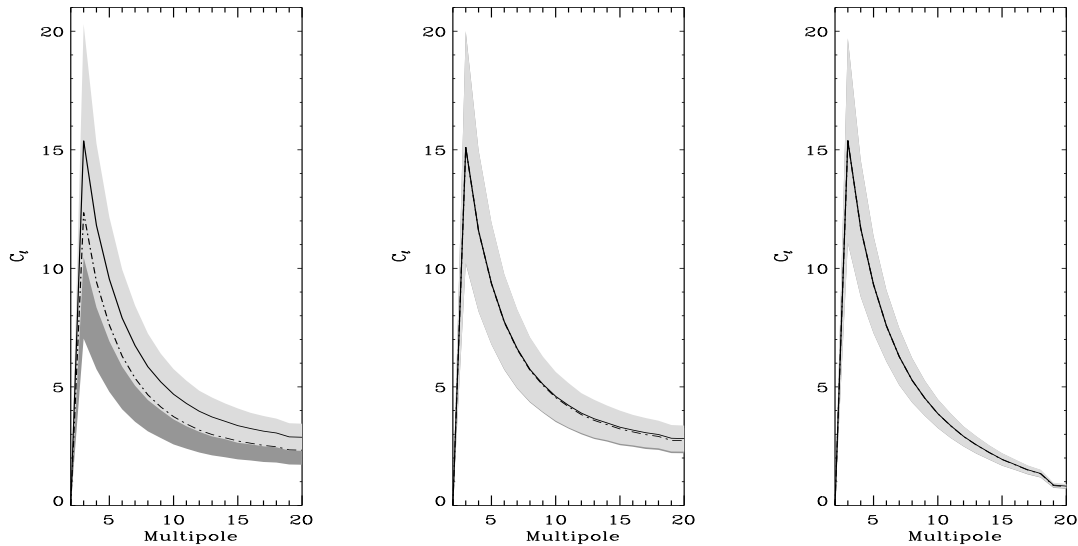}
\end{center}
\caption{Reconstructed CMB angular power spectrum from our $\sim 3000$ 
Monte Carlo (MC) realisations of fake CMB skies and instrumental noise. 
Left panel shows results from nominal noise case, middle panel is for uniform noise
with nominal mean amplitude and right panel is for uniform noise with reduced amplitude.
Solid lines refer to 53 and 90~GHz combination while dashed lines include the 31.5~GHz
in the analysis. Light grey shaded area refers to 1-sigma limit for 53 and 90~GHz channels
while dark grey is 1-sigma limit with also the 31.5~GHz. Those shaded areas
nearly fully overlap in all cases but the one with nominal noise.
Including the 31.5~GHz channel in the analysis leads to power spectra that are suppressed
with respect to the case in which it is excluded. Small deviations are visible also for
the uniform noise and mean nominal amplitude case.}
\label{MCcmbPS}
\end{figure*}
One of the assumptions of \fstica\ is that the noise is Gaussian, stationary
and uniformly distributed on the sky. The latter is manifestly not
true for DMR. 
In order to validate the impact of noise amplitude and 
distribution on the reconstruction procedure,
we performed $\sim 3000$ simulations of fake CMB skies (characterised
by a Harrison-Zeldovich initial power law spectrum with an rms
quadrupole normalisation of $\sim\ 18 \mu$K) and instrumental noise.
Specifically we consider three cases:
\begin{enumerate}
\item DMR nominal noise amplitude and distribution 
\item uniform noise distribution (i.e. each pixel is observed the same number of times) 
	with the noise amplitude matched to the mean noise amplitude
        in the actual DMR data (0.322, 0.100 and
	0.139 mK at 31.5, 53 and 90~GHz, respectively)
\item uniform noise distribution with amplitude reduced by a factor of 4 with respect
	the nominal values
\end{enumerate}

For each realisation we performed a component separation on the cut sky 
with the {\em p}, {\em g} and {\em t} functions. We have considered
the cases where all three frequencies are included, or just the 90 and
53~GHz data (since we know that
the 31.5~GHz channel shows a different noise pattern on the sky). In each case
we evaluate the frequency scaling (between 90 and 53~GHz) of the reconstructed 
CMB component as well as its angular power spectrum on the cut sky
exploiting the technique described in G\'orski \etal~(1994) and in 
G\'orski \etal~(1996, hereafter G96).

\begin{table}
\begin{center}
\caption{Reconstructed CMB frequency scaling between 90 and 53 GHz for our 3000
MC simulations for three different models of the noise distribution as explained
in the text. There is a clear degradation of the result when the 31.5~GHz
channel is included in the analysis.}
\label{MChistoscal}
\begin{tabular}{lccc}
\hline\hline
\ Combination & i & ii & iii \\
\hline
\ 31.5:53:90 & 0.871$\pm$0.477 & 0.879$\pm$0.188 & 0.876$\pm$0.019 \\
\ 53:90      & 0.871$\pm$0.205 & 0.878$\pm$0.192 & 0.876$\pm$0.019 \\
\hline
\end{tabular}
\end{center}
\end{table}
Table~\ref{MChistoscal} reports the results for the derived frequency scaling for
the {\em p} function (similar results are obtained with {\em g} and {\em t}).
The results are almost identical for noise cases ii) and iii) with
either 2 and 3 DMR frequencies, though naturally, with 
higher sensitivity a higher accuracy is achieved for
the derived CMB frequency scaling.
However, the situation is quite different for the nominal noise
distribution and amplitude
(case i). Although the mean value is almost the same and consistent with the
expected theoretical value, the rms is somewhat degraded being
0.477 and 0.205 for 3 and 2 frequency channels separation, respectively.
Therefore a first indication of the relevance of the noise distribution is derived from
an inspection of the frequency scalings.

A similar situation is found in the analysis of the reconstructed 
CMB component power spectrum as shown in Figure~\ref{MCcmbPS}.
The CMB component recovered when all three DMR sum maps 
are inputs to the \fstica\ algorithm shows an angular 
power spectrum (dot-dashed line) that is suppressed with respect to that derived
from only 90 and 53~GHz inputs (solid line), although they do agree at the 1-sigma level
(dark grey and light grey shaded regions for all DMR frequencies and only
90 and 53~GHz respectively).
The reason for this behaviour is probably associated with the distinct noise distribution 
of the 31.5~GHz channel 
relative to the other two frequencies, which have considerably more consistent noise patterns.
This drives the \fstica\ algorithm to identify two components:
one which has the expected CMB frequency scaling
and a second with an unphysical (negative) frequency scaling 
(and interestingly a higher $S/N$ ratio) into which
part of the CMB power has been aliased.

We conclude that differences in the noise patterns
between the DMR channels are a potential source of unphysical components reconstructed 
with \fstica, and can have a notable effect on the frequency scaling of the 
reconstructed CMB component and its angular power spectrum.
Therefore in what follows we restrict analysis to the 90 and 53~GHz channels only.

\begin{figure*}
\begin{center}
\includegraphics[height=4.in,angle=90]{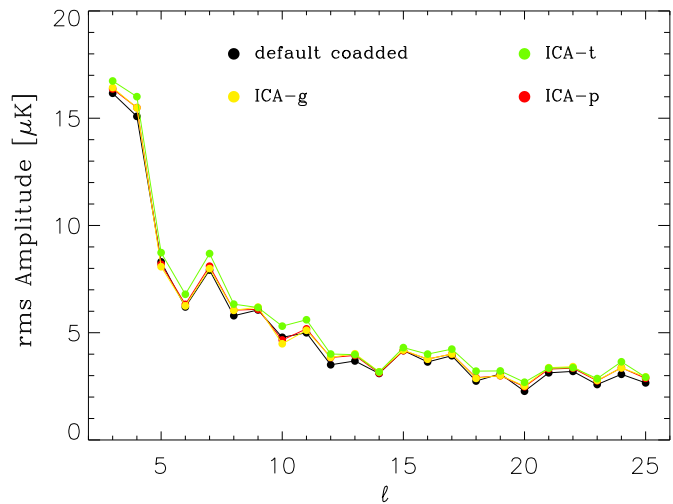}
\end{center}
\caption{Angular power spectra from the 90 and 53~GHz combination of DMR data (default
coadded) and from the reconstructed CMB ICA components with {\em p-}, {\em g-} and {\em t-}functions.}
\label{DMRcmbPS}
\end{figure*}
\begin{figure*}
\begin{center}
\vspace{-2cm}
\hspace{1cm}
\includegraphics[width=3.in,angle=90]{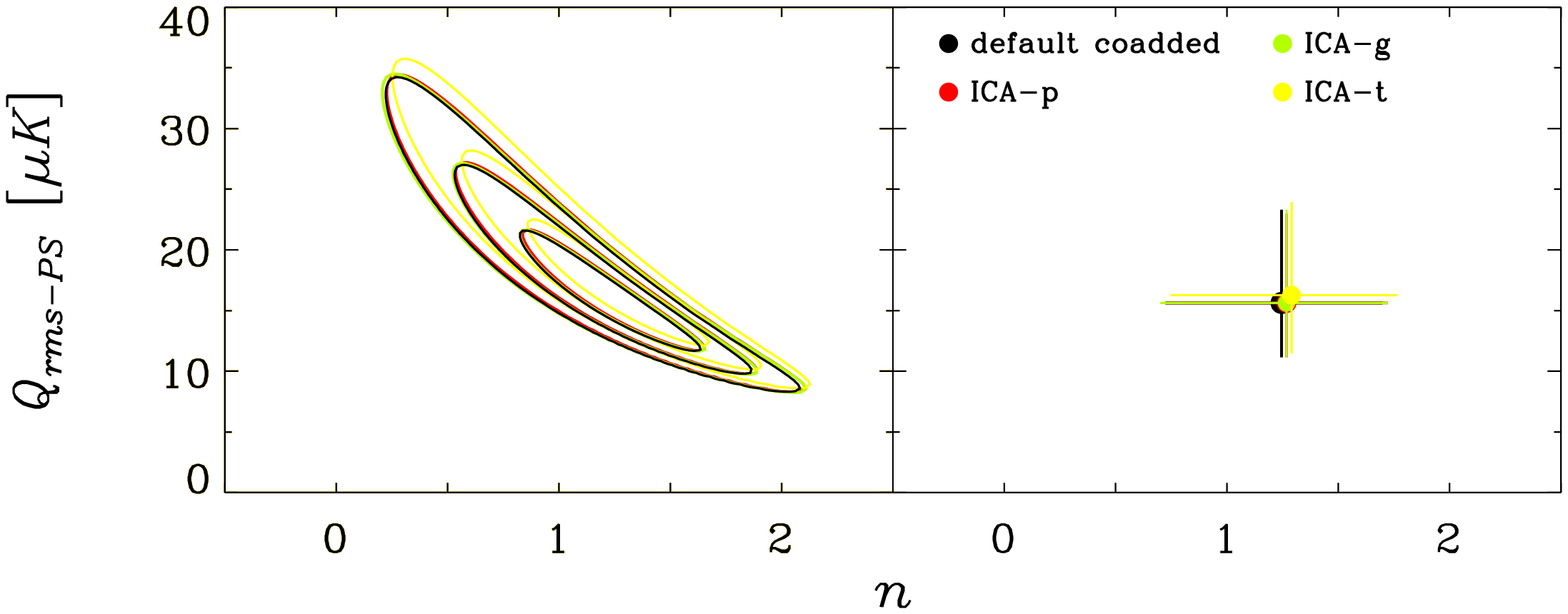}
\vspace{-2.cm}
\end{center}
\caption{Likelihood contours in the \qrms\ ,\n\ plane for
the reconstructed CMB component with \fstica. Also plotted is the 
``classic'' result for the minimum noise combination of the 90 and 53~GHz
DMR channels.}
\label{like}
\end{figure*}
\section{Results -- CMB}
\label{results-cmb}

We have applied \fstica\ to the 90 and 53~GHz frequency channels.
The results are almost identical for the three non-quadratic functions considered. 
Table~\ref{DMRscal} summarises the reconstructed CMB frequency
scalings and the percentage deviation from the theoretical CMB frequency scaling. Results from
the {\em p-} and {\em g-}functions are consistent in terms of the correct frequency
scaling and $S/N$ ratio. Results determined using the {\em t-}non-quadratic function
reveal a non-optimal frequency scaling for the CMB component. Nevertheless the correct
normalisation could be derived.

\begin{table}
\begin{center}
\caption{DMR 90-53~GHz reconstructed CMB frequency scaling and $S/N$ ratio. Also
reported are the deviations from the theoretical expected CMB frequency scaling.}
\label{DMRscal}
\begin{tabular}{lccc}
\hline\hline
\ {\sc ICA} form & Scaling  & $\Delta$ & S/N\\
\hline
\ {\em p} & 0.82$\pm$0.28 & 6.6\% & 1.069 \\
\ {\em g} & 0.91$\pm$0.24 & 3.0\% & 1.065 \\
\ {\em t} & 0.64$\pm$0.23 & 30\%  & 1.072  \\
\hline
\end{tabular}
\end{center}
\end{table}

Figure~\ref{DMRcmbPS} shows the angular power spectrum of the reconstructed CMB
component for the {\em p-}, {\em g-} and {\em t-}functions compared
with that computed from the minimum-variance 
noise combination of the 53 and 90~GHz channels. All results are consistent
although the {\em t-}function shows a small excess of power on all
scales. At large $\ell$ this corresponds to a larger noise contribution.

Using the same linear combination coefficients that produce 
the reconstructed CMB components from the
input 90 and 53~GHz maps, it is possible to derive the rms noise
properties per pixel for the reconstructions. 
This noise prescription is used as an input
to the likelihood analysis (see G\'orski \etal~1994 and G96)
solving for \qrms\ and \n.
Table~\ref{fitDMR} reports the results of these fits, whilst
Figure~\ref{like} plots the corresponding likelihood contours.

\begin{table}
\begin{center}
\caption{\qrms\ and \n\ as derived from the CMB component
reconstructed with the \fstica\ algorithm.
Both 1- and 2-$\sigma$ confidence intervals and the corresponding 
results from G96 determined without galactic correction are reported.}
\label{fitDMR}
\begin{tabular}{ccc}
\hline\hline
\ Map & $Q_{\rm rms}$ & $n_{\rm s}$ \\
\hline
\ G96     & 15.63$^{+3.19}_{-2.56}$~$^{+7.67}_{-4.48}$ & 1.25$^{+0.22}_{-0.25}$~$^{+0.45}_{-0.53}$\\
\ & & \\
\ {\em p} & 15.63$^{+3.19}_{-2.56}$~$^{+7.35}_{-4.48}$ & 1.27$^{+0.25}_{-0.27}$~$^{+0.45}_{-0.55}$ \\
\ & & \\
\ {\em g} & 15.63$^{+3.19}_{-2.24}$~$^{+7.67}_{-4.48}$ & 1.27$^{+0.27}_{-0.27}$~$^{+0.45}_{-0.57}$ \\
\ & & \\
\ {\em t} & 16.27$^{+3.51}_{-2.56}$~$^{+7.67}_{-4.80}$ & 1.29$^{+0.25}_{-0.25}$~$^{+0.48}_{-0.54}$ \\
\hline
\end{tabular}
\end{center}
\end{table}

The results are consistent for {\em p} and {\em g} while
larger \qrms\ and \n\ values are recovered for {\em t}.
Nevertheless, all are consistent with the fit  
derived from the optimal minimum-variance combination
of the 90 and 53~GHz sky maps without foreground corrections 
applied. 

\section{Adding ``priors'' on foregrounds}
\label{fore}

In the previous section we detailed the results of our analysis
of \fstica\ as applied to the 53 and 90 GHz sky maps alone,
and stressed the consistency of the cosmological results with those
determined in G96 without the application of any galactic
foreground correction. The relevance of this comparison 
can  be understood as follows: with only two input sky maps,
the \fstica\ algorithm can only reconstruct two outputs,
and we have determined that these correspond to CMB together with
an unphysical, noise-related component, i.e. the method is not
able to reconstruct the galactic emission at high galactic latitudes.
Indeed, we have verified by means of simulations -- which include a 
galactic contribution modelled on a DIRBE 140\um\ template
(scaled to the DMR frequencies according to the correlation
coefficients from G96) -- that the $S/N$ ratio of the galactic emission
in the DMR data
is too low to allow \fstica\ to recover it.
Only when the noise rms is reduced by a factor of 10
relative to the actual DMR values can some galactic emission be separated by
\fstica, although the separation remains less than optimal.

In order to attempt to extract and
better separate the CMB signal from that foreground emission 
which is certainly present in the data,
we have proceeded to add additional channels (priors) using foreground templates
such as the Haslam map of the diffuse galactic radio emission at 408~MHz 
(Haslam \etal~1982) and the DIRBE map of galactic diffuse infra-red emission
at 140\um. When applying \fstica\ to the 53 and 90~GHz DMR data in combination with 
the Haslam and DIRBE templates,
we have assumed that the noise contribution to the template maps is small compared
to that in the DMR data, which is certainly the case.
Since we now work with 4 inputs, \fstica\ returns 4 derived components.
One is again almost consistent with noise ($S/N$ lower that 1), one has the CMB
frequency scaling and the expected $S/N$ ratio while the other two show features
respectively of the Haslam and DIRBE maps. It is important to note that \fstica\ does not
completely mix the Haslam and DIRBE signals into a single foreground component,
but is able to keep them largely separated (see \ref{results-fore}). 
As before, we fit the CMB component for \qrms\ and
\n. Table~\ref{galcorr} reports the results:
there is a general agreement between our results, especially those
determined with the {\em p-} and {\em g-}functions, and those reported in G96
after galactic emission has been taken into account using the template
Haslam and DIRBE maps.

\begin{table}
\begin{center}
\caption{\qrms\ and \n\ determined from an analysis of the CMB
component reconstructed by \fstica\ when additional foreground
channels (Haslam and DIRBE maps - HD)
are included. Both 1- and 2-$\sigma$
limits and the results from G96 
after galactic correction are reported.}
\label{galcorr}
\begin{tabular}{ccc}
\hline\hline
\ Map & $Q_{\rm rms}$ & $n_{\rm s}$ \\
\hline
\ G96+HD     & 14.99$^{+3.19}_{-1.92}$~$^{+7.35}_{-4.16}$ & 1.27$^{+0.20}_{-0.27}$~$^{+0.43}_{-0.55}$\\
\ & & \\
\ {\em p}+HD & 14.03$^{+2.56}_{-2.56}$~$^{+6.39}_{-4.16}$ & 1.27$^{+0.25}_{-0.27}$~$^{+0.47}_{-0.57}$\\
\ & & \\
\ {\em g}+HD & 14.67$^{+2.87}_{-2.24}$~$^{+7.03}_{-4.16}$ & 1.29$^{+0.20}_{-0.27}$~$^{+0.46}_{-0.54}$\\
\ & & \\
\ {\em t}+HD & 11.15$^{+2.24}_{-1.60}$~$^{+5.44}_{-3.19}$ & 1.34$^{+0.18}_{-0.25}$~$^{+0.47}_{-0.53}$\\
\hline
\end{tabular}
\end{center}
\end{table}

It is clear that {\em p} and {\em g} almost give the same
results while large differences are observed with the {\em t} form. 
This may be due to the low $S/N$ ratio of the DMR data: Maino \etal~(2002)
did not observe such behaviour when working with higher fidelity {\sc Planck}
simulations. Inspection of the extracted components shows that the frequency
scalings of the CMB component are almost the same as when only the DMR data 
were used as inputs (variations are $\lsim 1\%$) and that the extracted signals have also the same
$S/N$ ratio. We also computed the final rms in the extracted maps and compare them
with the previous DMR-only case. The rms figures are almost the same for {\em p} and {\em g}
(deviations within 8\%) while for the {\em t} form we observe a variation in the map
rms of about 30\%. This implies that the {\em t} is unable to
adequately separate the underlying components and identify them as CMB and galactic signals.

\section{Results -- Foreground emissions}
\label{results-fore}

\begin{figure*}
\begin{center}
\includegraphics[width=5cm,angle=90]{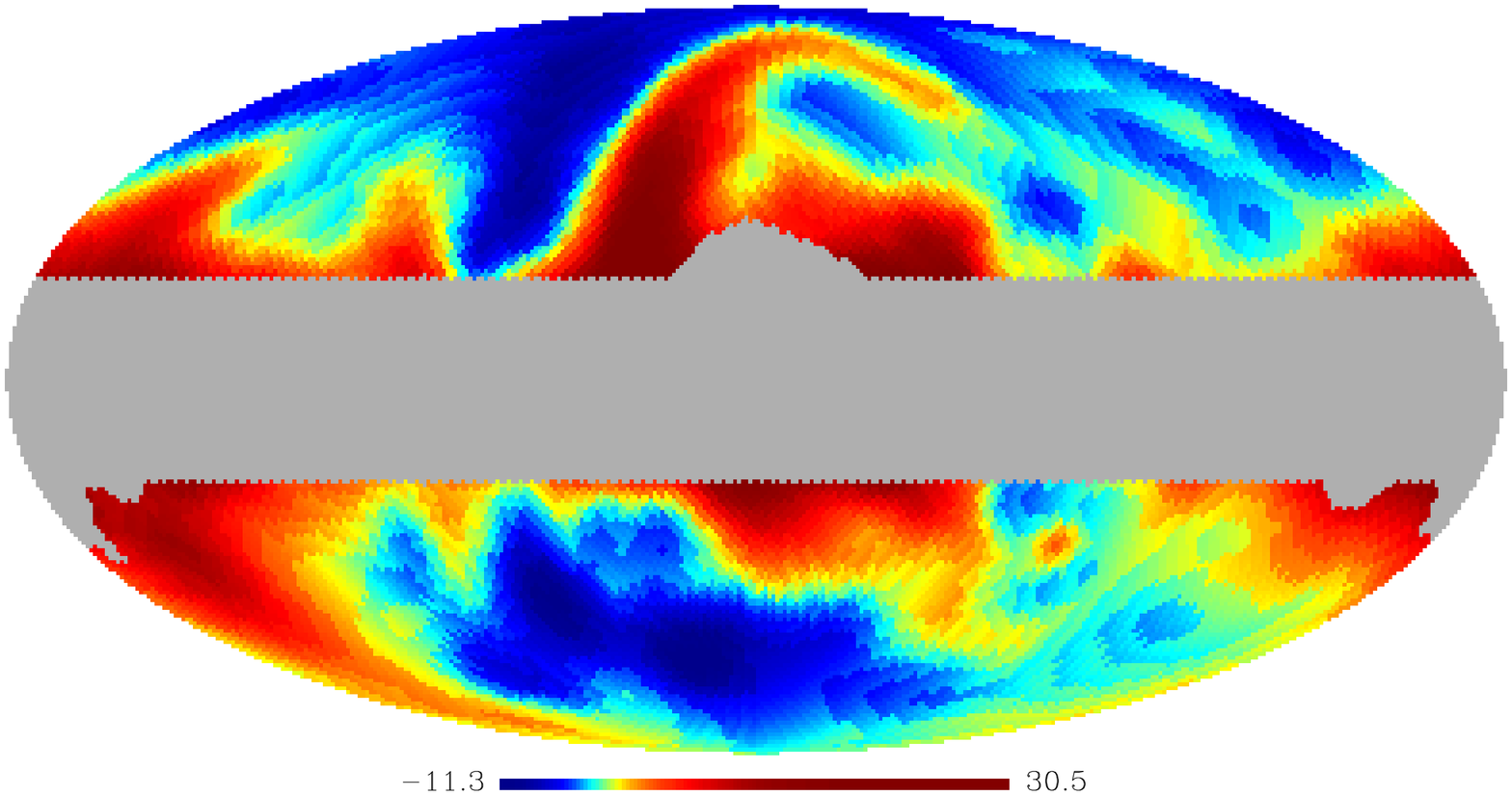}
\includegraphics[width=5cm,angle=90]{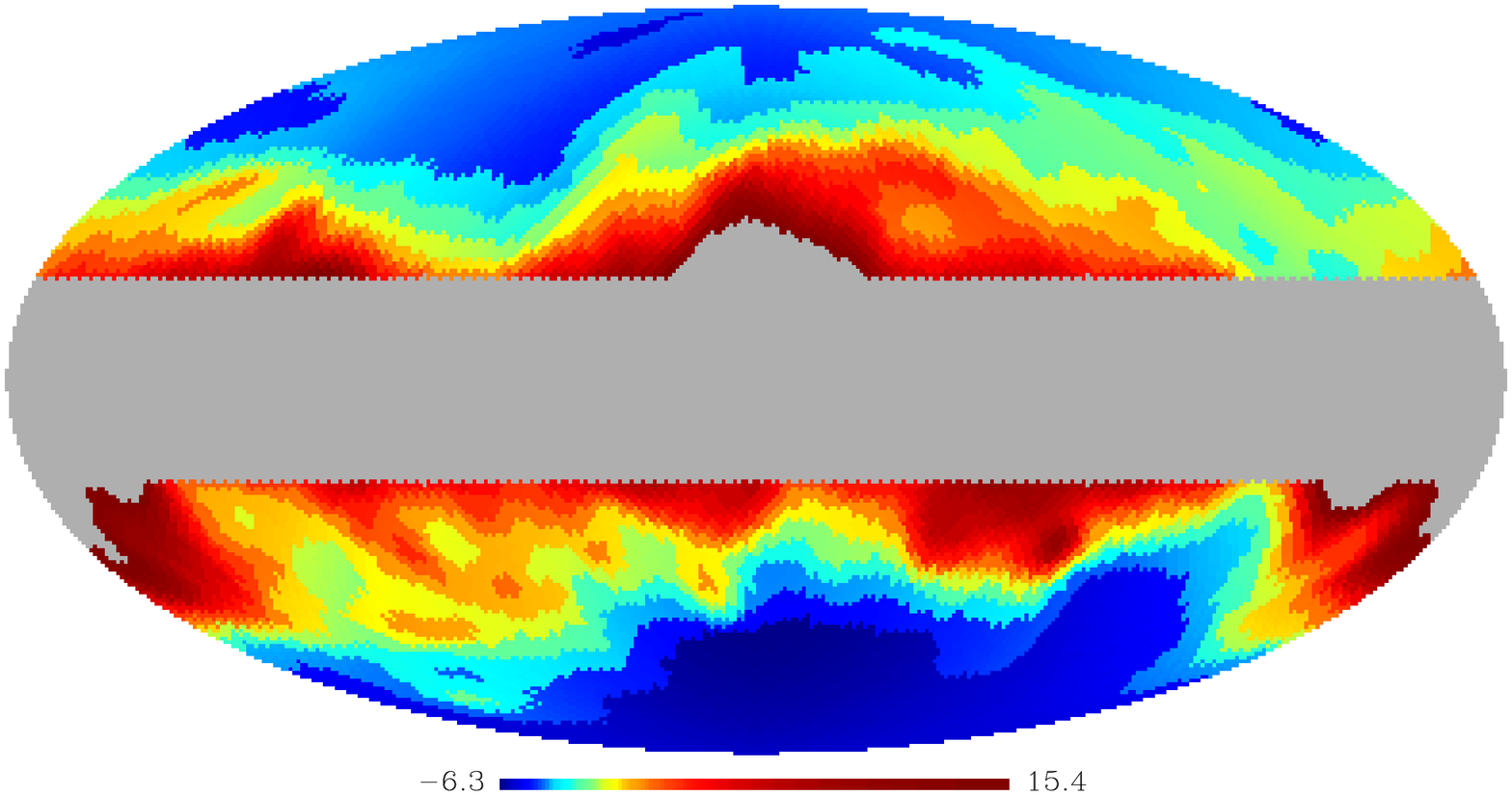}
\includegraphics[width=5cm,angle=90]{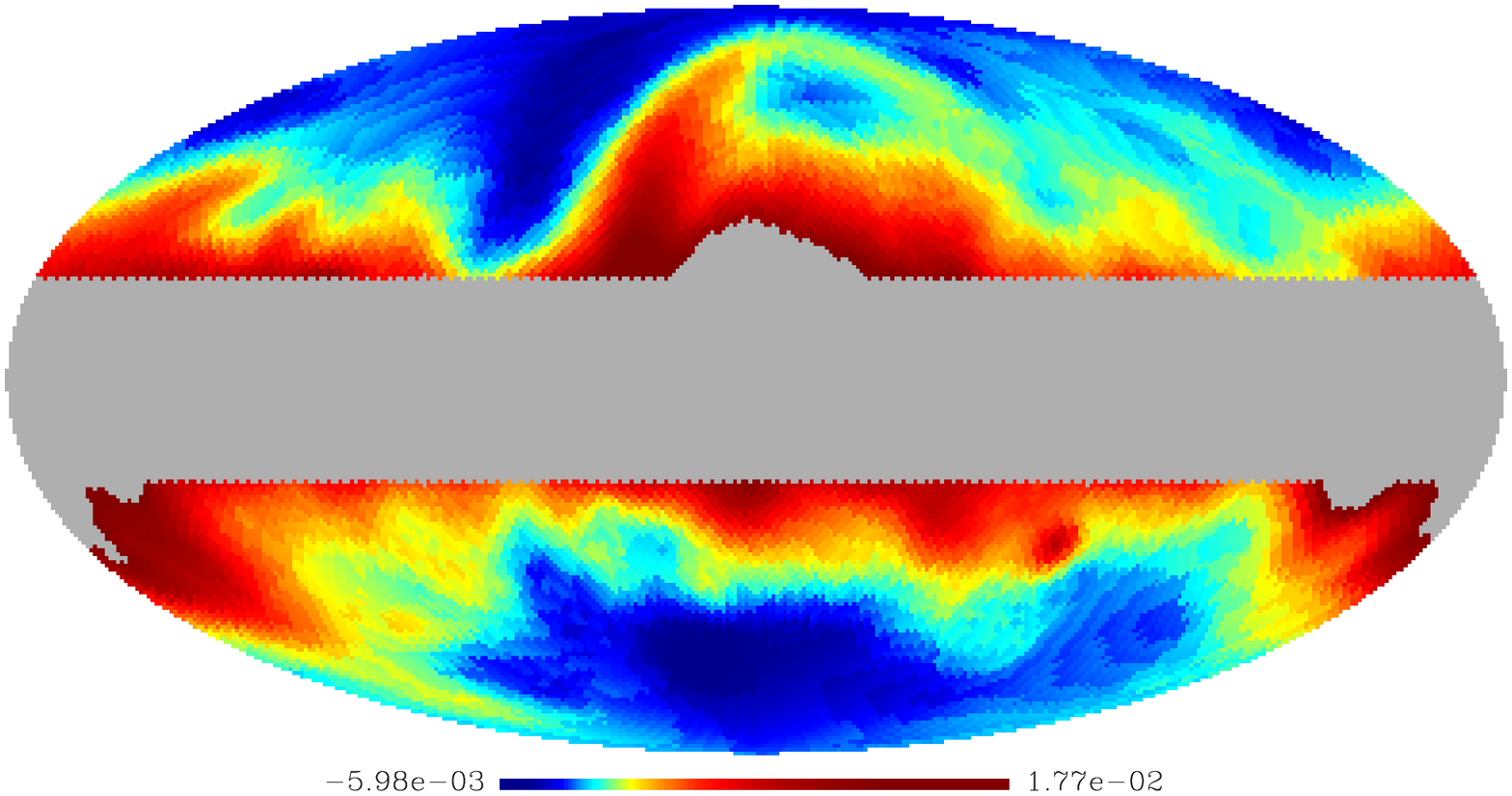}
\includegraphics[width=5cm,angle=90]{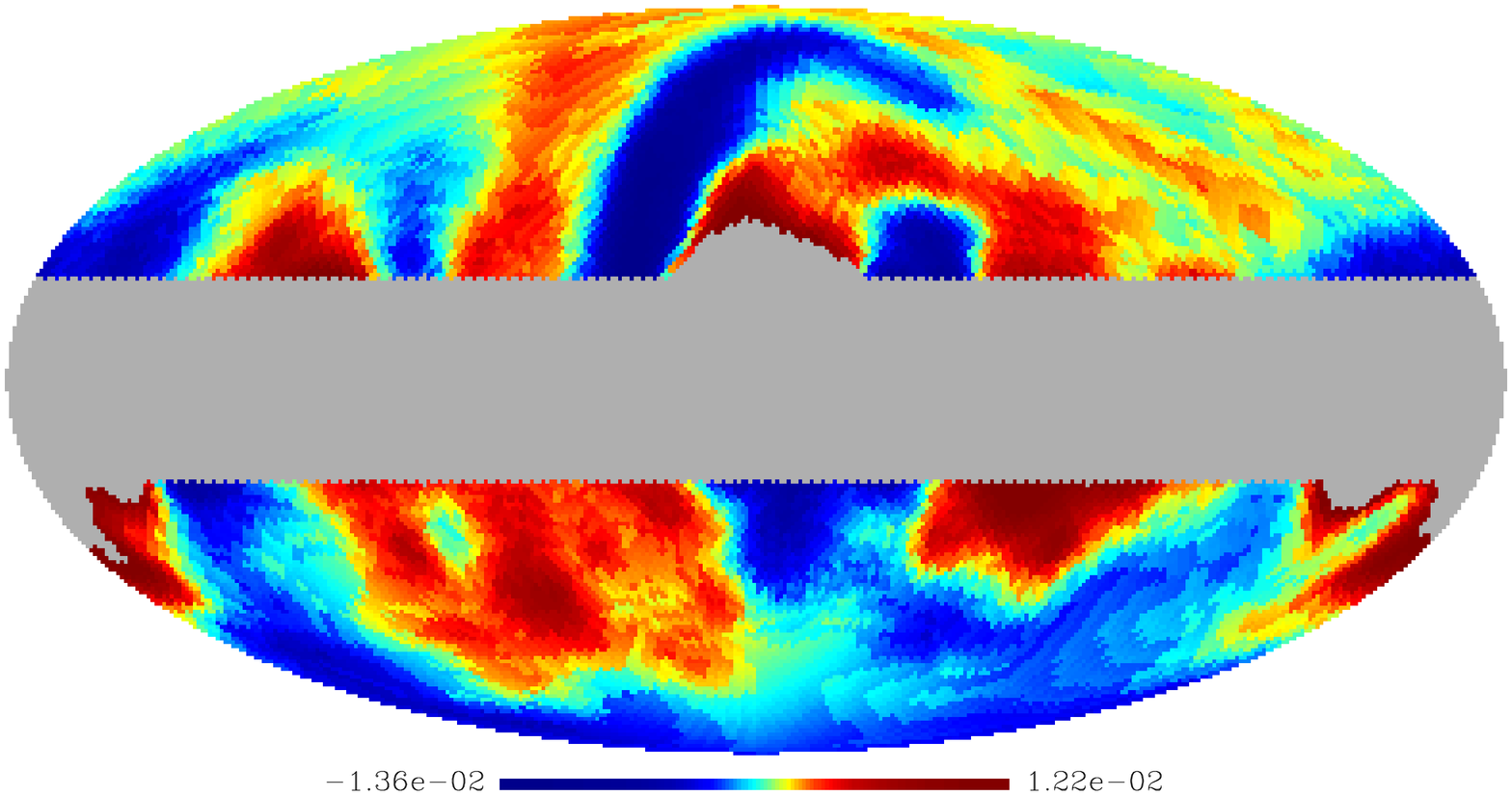}
\end{center}
\caption{Top panels show the input Haslam 408 MHz (left) and DIRBE 140\um\ (right) 
foreground maps
while bottom panels report the reconstructed ICA component maps derived with
the {\em t}-non-quadratic form. The ICA components show
large noise and signal ghosts from the other component (e.g. the synchrotron North
Polar Spur is visible with a negative amplitude in the reconstructed dust map
and high latitude structures in the DIRBE map are aliased into the reconstructed
synchrotron map). Units for Haslam and DIRBE maps are K and MJy/sr respectively
and are the same for the ICA components (normalized at 90~GHz).}
\label{ghost}
\end{figure*}

We have analysed the two additional signal components extracted by \fstica.
These appear to closely resemble the Haslam and DIRBE maps, 
but this is not optimal (the Spearman correlation coefficient $r_{\rm S}\sim 90\%$).
Furthermore, visual inspection reveals
that some structure which appears in the input Haslam map template
appears as ``ghosts'' (with negative signal) 
in the reconstructed dust map, and viceversa as shown in Figure~\ref{ghost}.
Such behaviour almost certainly reflects a more complex 
spectral signature of the foreground components than the method
(with only 4 input data sets) is able to disentangle. Undoubtedly,
the significant large angular scale spatial correlation between foregrounds in various
regions of the sky and lower fidelity of the foregrounds due to both 
the $S/N$ and large angular smoothing of the DMR data also play a role.
In fact, the extent of the signal aliasing depends on the non-quadratic form 
considered in the analysis, and is particularly evident for the {\em t} form. 
The spectral indices for these galactic independent components derived
are not robust to the choice of non-quadratic form
and furthermore are not consistent with what expected from previous studies (e.g. G96). 
However, the CMB component is reconstructed well and remains clean from galactic
emission as results in Table~\ref{galcorr} show. 

Interestingly, this means that \fstica\ {\em can} provide information on foregrounds, 
not directly from the (spectral) properties of the reconstructed 
galactic components, but rather from information gleaned in the reconstruction
of the CMB component.
Each ICA component is obtained by a weighted {\em linear} combination of the
input data, even though the combination coefficients (actually the rows of the
matrix $\mathbf{W}$) are computed by a non-linear process. 
This means that we can interpret the derived weights 
effectively as correlation coefficients between the foreground templates
and the DMR data.
Indeed, for the previous analysis with the Haslam and DIRBE maps,
the recovered values are similar to that found in G96 using an 
explicit (minimum $\chi^2$) correlation analysis. 
Furthermore, the consistency of the results between {\em p}, {\em g} and, to
some extent, {\em t}, suggest that the method allows a robust estimation of the
correlation of the DMR data with the considered templates.

Recently Banday \etal~(2003) have derived correlation
coefficients between the DMR data and three foreground templates: 
the Haslam map at 408~MHz, the DIRBE map at 140\um\ and
a new map of \ha\ emission (Dickinson, Davies \& Davis 2003)
that could be used to determine a reliable free-free template.
Furthermore when thermal dust and free-free
emission are properly accounted for, they found a strong
anomalous correlation between data at 31.5~GHz and the DIRBE map.
This component is found to be well fitted by a spectral
shape of the form $\nu^{-\beta}$ with $\beta \sim 2.5$.
 
We attempt to verify here whether \fstica\ is able to
arrive at similar results, and we therefore extend our analysis 
to include the \ha\ template. 
An initial consideration is that the non-quadratic
forms of the \fstica\ algorithm are in general sensitive to different
features of the astrophysical components, therefore it is possible 
that a given form may or may not be particularly well matched to the detection
of certain signals on the sky.
We therefore calibrate which form is best suited
to our current studies, by using results from Banday \etal~(2003)
to construct a fake skies with CMB, foreground emissions (whose
correlation coefficients are known a-priori) and $\sim 10000$ 
realisations of the DMR instrumental noise. 
In each run a single DMR channel is considered together
with foreground emission templates and each of
{\em p}, {\em g} and {\em t} is employed to evaluate
the correlation coefficients. These simulations also provide
an indication of the uncertainty in the reconstruction process.
Simulations show that {\em p} and {\em g} provide
consistent results.

Table~\ref{corr} reports our estimates of the correlation 
coefficients at each of the DMR frequencies.
\begin{table}
\begin{center}
\caption{\fstica\ correlation coefficients between each of the
DMR frequencies and foreground templates. Results are reported for the
best calibrated non-quadratic form in each case. Errors are
derived from simulations. Units are
$\mu$K $X^{-1}$, where $X$ are template units.}
\label{corr}
\begin{tabular}{lccc}
\hline \hline
\ Component & \multicolumn{3}{c}{DMR Map (GHz)} \\ \cline{2-4} 
\ & 31.5 & 53 & 90 \\
\hline
\ Dust        & 5.19$\pm$1.55 &  2.58$\pm$1.09 & 1.91$\pm$1.14 \\
\ Synchrotron & 1.77$\pm$0.90 & -0.08$\pm$0.67 &-0.35$\pm$0.61 \\
\ Free-free   & 3.19$\pm$1.48 &  0.86$\pm$0.68 & 0.98$\pm$0.86 \\
\hline
\end{tabular}
\end{center}
\end{table}
\begin{figure}
\begin{center}
\includegraphics[width=3.5in]{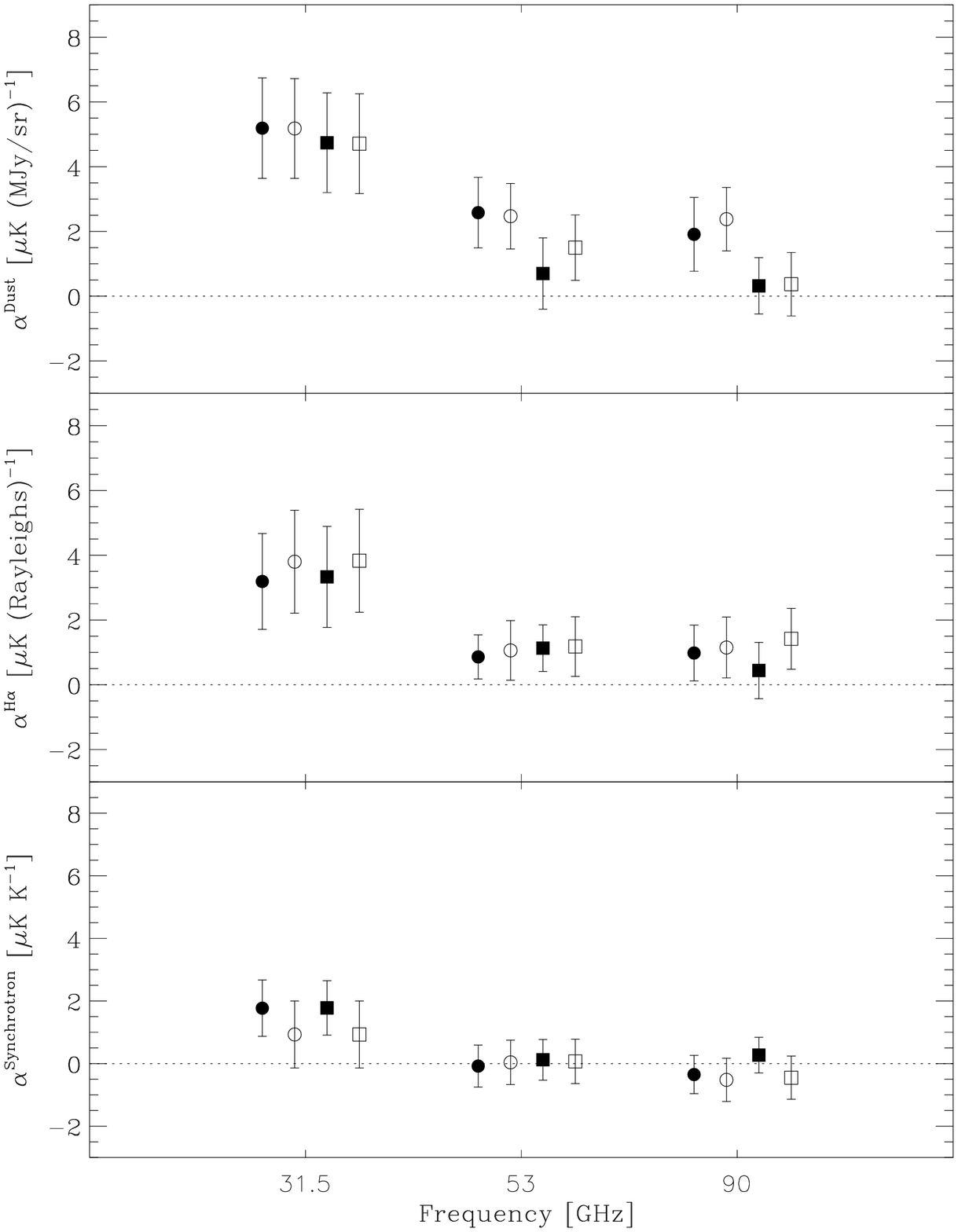}
\end{center}
\vspace{.3cm}
\caption{Derived correlation coefficients between DMR at 31.5, 53 and 90~GHz
and the three foreground emission templates. Dust emission is traced
by DIRBE map at 140\um, synchrotron emission by Haslam
map at 408~MHz and free-free emission by \ha\ map. Filled symbols are
results obtained with \fstica\ while open symbols are from
Banday \etal~(2003) using cross-correlation technique. Circles refer to
test without subtraction of dust thermal model while squares are 
when model 7 of FDS has been subtracted from DMR data.}
\label{mainobanday}
\end{figure}
Figure~\ref{mainobanday} plots our results (filled circles) together
with those from Banday \etal~(2003) (open circles). 
There is good general agreement between the two sets of results.
\fstica\ also clearly detects foreground emission correlated
with the DIRBE dust map at 140\um\ and identifies this as the major foreground
emission at 31.5~GHz. 

In order to further probe the nature of this anomalous 
dust correlated component, we subtract from the DMR
data model 7 of thermal dust emission from Finkbeiner, Davis
and Schlegel~(1999, hereafter FDS). This comprises two dust components,
one at a temperature of 9.6~K with dust emissivity $\alpha = 1.5$ and the other at 
16.4~K with $\alpha = 2.6$. The results are shown in 
Table~\ref{corrnodust}.

\begin{table}
\begin{center}
\caption{\fstica\, correlation coefficients between each of the
DMR frequencies, after thermal dust emission removal
following model 7 in FDS, and foreground templates. Results are reported for the
better ``calibrated'' non-quadratic form in each case. Errors are
derived from simulations. Units are
$\mu$K $X^{-1}$, where $X$ are template units.}
\label{corrnodust}
\begin{tabular}{lccc}
\hline \hline
\ Component & \multicolumn{3}{c}{DMR Map (GHz)} \\ \cline{2-4} 
\ & 31.5 & 53 & 90 \\
\hline
\ Dust        & 4.74$\pm$1.54 & 0.70$\pm$1.10 & 0.32$\pm$0.87 \\
\ Synchrotron & 1.78$\pm$0.87 & 0.12$\pm$0.65 & 0.27$\pm$0.57 \\
\ Free-free   & 3.33$\pm$1.56 & 1.13$\pm$0.72 & 0.44$\pm$0.87 \\
\hline
\end{tabular}
\end{center}
\end{table}

From Table~\ref{corrnodust} and Fig.~\ref{mainobanday} it is clear
that, if FDS model 7 provides an accurate prediction for the thermal
dust emission at DMR frequencies, then it must account for essentially
the entire dust correlation at 90~GHz. At 31.5~GHz a strong anomalous
correlation with the DIRBE 140\um\ dust template persists 
after removal of the thermal dust model, whilst the results at 53~GHz
seem to indicate the presence of both a thermal and anomalous 
dust correlated component which would have roughly 
equal contributions at a frequency of $\sim$ 60~GHz. 

We investigate the spectral behaviour of this anomalous dust
correlated component by fitting our data points (the correlations
coefficients with DIRBE 140\um\ after FDS model 7
subtraction) by a power law of the form 
$\mathbf{A}_{\rm norm}(\nu/\nu_0)^{-\beta}$.
$\mathbf{A}_{\rm norm}$ has units of $\mu$K/(MJy sr$^{-1}$)
and is the amplitude of the
emission at the reference frequency $\nu_0$ of 31.5~GHz. 
Table~\ref{fit} shows the results from this analysis:
it is clear that the frequency behaviour is consistent with a spectrum steeper
that the typical value of $\sim$ 2.15 expected for pure free-free emission 
(although they are still compatible with this kind of
emission). 

\begin{table}
\begin{center}
\caption{Power law spectral indices from pairs of
frequencies and from a fit using data from 31.5 to 90~GHz.
We fit for $\mathbf{A}_{\rm norm}(\nu/\nu_0)^{-\beta}$
where $\nu_0$ is taken as 31.5~GHz. Also reported the
68\% confidence level errors.}
\label{fit}
\begin{tabular}{cc}
\hline\hline
\ $\beta_{31:53}$ & 3.67$\pm$2.01 \\
\ $\beta_{53:90}$ & 1.48$\pm$2.95 \\
\hline
Fitted $\beta$ & 2.56$^{+1.56}_{-0.98}$ \\
\hline
Fitted $\mathbf{A_{\rm norm}}$ & 3.90$^{+1.54}_{-0.99}$ \\
\hline
\end{tabular}
\end{center}
\end{table}

Such conclusions are consistent with those of by Banday \etal~(2003),
and are particularly satisfactory
due to the independence of the two techniques employed,
and the fact that the ICA method makes no assumptions
about the cosmological signal present in the data.

\section{Critical Discussion and Conclusion}
\label{conclusion}

In this paper we have applied for the first time the \fstica\ algorithm
to real CMB data and specifically to the \COBE\ 4-yr data.

One of the requirements for applicability of \fstica\ is that noise
has to be Gaussian and uniform distributed on the sky. Unfortunately,
this not the case for the DMR observing strategy. Furthermore part
of the original data at 31.5~GHz were removed to eliminate potential
contamination from a particular systematic effect (Kogut \etal~1996),
resulting in a noise distribution that is notably different
than that at 53 and 90~GHz.
Therefore, before applying the technique, we have tested
the effect of such a non-uniform noise distribution on both the frequency
scaling and power spectrum reconstruction of the recovered CMB component.
We run $\sim 3000$ MC simulations and considered three scenarios 
of different noise distribution in the data:
one nominal, one with uniform noise distribution and nominal mean amplitude
and the third with uniform distribution and reduced amplitude.
Results indicate that the different noise distribution of the 31.5~GHz 
channel results in the identification of a spurious non-physical component, 
and the degradation of the frequency scaling of the reconstructed 
CMB component. Furthermore, 
the angular power spectrum of the latter has suppressed amplitude with respect to the case
in which only the 90 and 53~GHz channels are considered,
resulting from the aliasing of CMB power into the non-physical component
associated with the noise asymmetry.

Further analysis was therefore restricted to include the 90 and 53~GHz channels only.
In this case we obtained a CMB reconstruction with a frequency scaling 
consistent with the expected theoretical one.
We fit the cosmological component with a power-law model parametererised
by the rms-quadrupole normalisation \qrms\ and spectral index \n, 
and find that the results are robust
against the choice of non-quadratic function, 
completely consistent with those 
derived in G96 from the optimal minimum-variance combination 
of the 90 and 53~GHz channels, without foreground corrections.
This is a very important outcome since equivalent results 
have been obtained with independent approaches.

Simulations with different noise rms properties have shown that foreground emission
in the DMR data has too low a $S/N$ ratio to be detected and separated with \fstica.
Only with a $S/N$ a factor of 10 or better does it become possible to 
identify foreground emission.
However, after adding two additional channels -- the Haslam map or radio emission
at 408~MHz and the DIRBE map of 140\um galactic infra-red emission --
\fstica\ is now able to
recognise foreground emission in the DMR data and separate it from 
CMB signal. The CMB frequency scaling for different \fstica\ forms are
consistent with that found without priors on foregrounds.
More importantly, we derived values for $Q_{\rm rms}$ and $n_{\rm s}$
for this new cleaned CMB reconstruction in very good agreement
with those reported in G96 after foreground correction.

The situation with the foreground component reconstruction is
less satisfying, however.
We found that the foreground spectral indices derived from
the ICA foreground components are not robust against the choice
of the \fstica\ non-quadratic form and this could be due to several reasons. First of all
the different forms tend to preferentially discriminate between different morphological
aspects of the foreground emission, with the residuals tending to be mixed together. 
Secondly, the $S/N$ ratio for foregrounds at 
high galactic latitudes is too low for \fstica\, to obtain faithful information
on these emissions. Finally, and quite realistically, it is likely that the true foreground
emission at microwave frequencies is different from our priors which are
based on observations at considerably different wavelengths.
It is worth noting, however, that Maino \etal~(2002) have demonstrated that, for 
higher $S/N$ observations and with an exact correspondence between ``priors'' and foreground
signals, the algorithm was able to properly extract the underlying
foreground contributions.

Despite the fact that the spectral behaviour of the derived foreground components 
is not properly recovered, our results do still allow some information on the 
non-CMB emission to be determined. Since the ICA components are linear 
combinations of the input sky maps, the weightings between the various
input data used to reconstruct the CMB sky can be interpreted as
correlation coefficients between the DMR data at specific frequencies 
and the foreground templates, in the spirit of the analysis in G96.
Indeed, the results in Table~\ref{galcorr} are in substantial agreement
with this. Following the recent work of Banday \etal~(2003)
we have further considered an extra channel using the \ha\ template provided
by Dickinson, Davies \& Davis (2002). Using these three templates, \fstica\ 
is able to derive the correlation coefficients between the DMR data and
foreground templates in very good agreement with their results.
We found that after subtraction of the dust thermal emission
as accounted for by model 7 of FDS, a strong dust correlated anomalous
component remains, and consitutes the strongest foreground emission at 31.5~GHz.
This component is found to be well fitted by a power law spectrum
with $\beta \sim 2.56$.

Blind component separation algorithms such as \fstica\, are promising and
can be applied to real data even if their assumptions are not completely
satisfied. 
The results obtained so far are encouraging and future application to
CMB datasets (e.g. $WMAP$ and {\sc Planck}) are foreseen.

\section*{Acknowledgements}
It is a pleasure to thank E.~Salerno for useful discussion. DM thanks A.J.~Banday for
warm hospitality at MPA and AJB thanks D.~Maino for kind hospitality at OAT in Trieste.
We thank Clive Dickinson for providing us with the \ha\ template prior to
publication.

\label{lastpage}
\end{document}